\documentclass[aps,pre,twocolumn]{revtex4}

\usepackage{epsfig,graphicx}

\begin{document}

\title{Statistical features of freely decaying two-dimensional
hydrodynamic turbulence}
\author{A.N. Kudryavtsev$^{a,b}$, E.A. Kuznetsov$^{a,c,d}$\/\thanks{kuznetso@itp.ac.ru},
 E.V. Sereshchenko$^{a,b,e}$}
\affiliation{\small \textit{$^{a}$ Novosibirsk State University,  630090 Novosibirsk, Russia} \\
\small \textit{ $^{b}$ Khristianovich Institute of Theoretical and Applied Mechanics, SB RAS, 630090 Novosibirsk, Russia}\\
\small \textit{$^{c}$ Lebedev Physical Institute, RAS, 119991 Moscow, Russia}\\
\small \textit{$^{d}$ Landau Institute for Theoretical Physics, RAS, 119334 Moscow, Russia}\\
\small \textit{$^{e}$ Far-Eastern Federal University, 690091 Vladivostok, Russia
}}

\begin{abstract}
Statistical characteristics of freely decaying two-dimensional
hydrodynamic turbulence at high Reynolds numbers are numerically studied.
In particular, numerical experiments (with resolution up to
$8192\times 8192$) provide a Kraichnan-type turbulence spectrum
$E_k\sim k^{-3}$. By means of spatial filtration, it is found that
the main contribution to the spectrum comes from the sharp vorticity gradients
 in the form of quasi-shocks. Such quasi-singularities
are responsible for a strong angular dependence of the spectrum owing
to well-localized (in terms of the angle) jets with minor and/or large
overlapping. In each jet, the spectrum decreases as $k^{-3}$.
The behavior of the third-order structure function accurately
agrees with Kraichnan direct cascade concept corresponding to a constant
enstrophy flux. It is shown that the power law exponents $\zeta_n$ for higher
structure functions grow more slowly than the linear dependence of $n$,
which testifies to turbulence intermittency.
\end{abstract}

\maketitle

\vspace{0.5 cm}

\vspace{0.5 cm}{\bf 1. Introduction.}
It is known that  developed two-dimensional hydrodynamic turbulence,
in contrast to three-dimensional turbulence, in the inertial interval
of scales possesses an additional integral of motion --- enstrophy.
Enstrophy is an integral of the squared vorticity $\int_{s}\Omega
^{2}d\mathbf{r}$. As Kraichnan demonstrated in 1967 \cite{kraichnan},
the existence of this integral generates in the inertial interval
its own Kolmogorov spectrum of turbulence:
\begin{equation}
E(k)~\sim ~k^{-3},  \label{Kraichnan}
\end{equation}
(now called the Kraichnan spectrum). This spectrum corresponds to a constant
enstrophy flux toward the region of small scales. Simultaneously, according to
\cite{kraichnan}, a conventional Kolmogorov spectrum $E(k)~\sim ~k^{-5/3}$
with a constant  energy flux directed toward the region of small values of
$k$ (inverse cascade) is formed in the region of large scales.
After that paper was published, many numerical experiments were
performed  (we note only some of them: \cite{lilly}-\cite{okhitani};
a more complete list can be found in \cite{KNNR-07}), which testified
in favor of existence of the Kraichnan spectrum. At the same time,
 already in the first numerical experiments there was observed
(see, e.g., \cite{lilly}) the emergence of sharp vorticity gradients (consistent with the high  Reynolds number). It corresponds
to the formation of jumps similar to shock waves which thickness is small
as compared with their length. Based on this observation, Saffman
\cite{saffman} proposed another spectrum, which differs from the Kraichnan
distribution:
$E(k)~\sim ~k^{-4}$.
The Saffman spectrum was obtained under the assumption that the main
contribution to the spectrum comes from isotropically distributed vorticity
shocks. On the other hand, it follows from simple considerations that the
spectrum with such shocks is expected to have the Kraichnan type behavior rather
than the Saffman distribution. Indeed, if one considers the Fourier transform
$\Omega _{k}$ of a vorticity jump $\Omega \propto \theta (x)$,
where $\theta (x)$ is the Heaviside function, then one can easily obtain
$\Omega _{k} \propto k^{-1}$, that immediately
yields the spectrum $E(k)~\sim ~k^{-3}$. However, the situation is not
too simple. If one assumes that the characteristic length of the step
$L\gg k^{-1}$, then the energy distribution from one such step
in the $k$-space has the form of a jet with an apex angle of the order of
$\left( kL\right) ^{-1}$. The maximum value of the energy distribution along
the jet decreases as $\propto k^{-3}$ \cite{KNNR-07,K-04,KNNR-10}.
As was demonstrated in \cite{KNNR-07,K-04}, such an estimate at $kL\gg 1$
follows from the rigorous analysis.

The main attention in this work is paid to the numerical study of statistical
characteristics of the direct cascade of freely decaying two-dimensional
turbulence and the influence of vorticity quasi-shocks on these
characteristics. As compared to previous works,

(i) the spatial resolution is significantly increased (up to $K\times K = 8192^{2}$
points). High-resolution numerical simulations confirmed all previous
results \cite{KNNR-07,KNNR-10};

(ii) it is proved numerically that the spectrum (\ref{Kraichnan}) arises
owing to vorticity quasi-shocks, which form a system of jets with weak
and strong overlapping in the $k$-space;

(iii) in each of these jets,  the decrease in $E(k)$ at high values of $k$
is proportional to $~k^{-3}$, which yields the spectrum (\ref{Kraichnan})
after averaging over the angle;

(iv) the third-order structure function $\langle (\delta\Omega)^2 \delta v_{\|}\rangle$
depends linearly on $R=|\mathbf{r}-\mathbf{r}'|$
(where $\delta\Omega=\Omega(\mathbf{r})-\Omega(\mathbf{r}')$, $\delta v_{\|}$
is the projection of the velocity difference onto the vector $\mathbf{R}$),
in complete agreement with the Kraichnan theory. The higher
structure functions of velocity have the exponents $\zeta _{n}$, which grow
more slowly than the first power of $n$, testifying to turbulence intermittency.

\vspace{0.5cm}{\bf 2. Main equations and numerical scheme.}
As was noted in Introduction, the emergence of sharp gradients of vorticity
was observed in the majority of numerical experiments (see, e.g.,
\cite{lilly}-\cite{okhitani}) modeling two-dimensional turbulence at high
Reynolds numbers, i.e., in the regime where the Euler equation can be considered
in the zero approximation instead of the Navier--Stokes equation.

In terms of vorticity $\Omega ={\partial v_{y}}/{\partial x}-
{\partial v_{x}}/{\partial y}$, the Euler equation is written as
\begin{equation}
\frac{\partial \Omega }{\partial t}+(\mathbf{v}\nabla )\Omega =0\quad \quad
\mbox {\rm with}\quad \quad \mbox{div}~\mathbf{v}=0,  \label{vort}
\end{equation}
i.e., $\Omega $ is a Lagrangian invariant. As was demonstrated in
\cite{KNNR-07,K-04}, the reason for the emergence of strong gradients of
vorticity in two-dimensional turbulence is connected with compressibility
of the field $\mathbf{B}=\mbox{rot}\,(\Omega \hat z$), which was called
divorticity. This vector is  tangent  to the level
lines $\Omega (\mathbf{r})=\mbox {\rm const}$, and the field $\mathbf{B}$
is frozen into the fluid, i.e., it satisfies the equation \cite{Sulem, weiss}
\begin{equation}  \label{B}
\frac{\partial \mathbf{B}}{\partial t} = \mbox {\rm
rot}~[\mathbf{v}\times\mathbf{B}].
\end{equation}
It follows from (\ref{B}) that $\mathbf{B}$ changes owing to the velocity
component $\mathbf{v}_{n}$ normal to $\mathbf{B}$, which is the velocity of the
lines of the field $\mathbf{B}$ due to its frozenness. The Lagrangian
trajectories determined by $\mathbf{v}_{n}$,
\begin{equation}
\frac{d\mathbf{r}}{dt}=\mathbf{v}_{n}(\mathbf{r},t);\quad
\mathbf{r}|_{t=0}=\mathbf{a},
\end{equation}
define the mapping
\begin{equation}  \label{map}
\mathbf{r}=\mathbf{r}(\mathbf{a},t).
\end{equation}
In this case Eq. (\ref{B}) admits integration \cite{KNNR-07,K-04}:
\begin{equation}  \label{BLR}
\mathbf{B}(\mathbf{r},t) = \frac{(\mathbf{B}_0(\mathbf{a})
\cdot\nabla_a)\mathbf{r}(\mathbf{a},t)}{J} ,
\end{equation}
where $\mathbf{B_0}(\mathbf{a})$ is the initial value of the field
$\mathbf{B}$, and $J$ is the Jacobian of mapping (\ref{map}).
As the trajectories (\ref{map}) are defined by the normal component
of velocity rather than by velocity proper, the Jacobian $J$
is not fixed. It can take arbitrary values, in particular,
it can tend to zero. This implies compressibility of the field
$\mathbf{B}$. It should be noted that Eq. (\ref{BLR}) is an
analog of the vortex lines representation, which was introduced in
\cite{KR,kuz} for the three-dimensional Euler equation.

It is known that the emergence of discontinuities in gas dynamics
is associated with vanishing of the Jacobian of the corresponding
compressible mapping, i.e., the transformation from the Eulerian to the Lagrangian
description. Just the decrease in $J$  is responsible for the
occurrence of sharp gradients of vorticity resulting from compressibility
$\mathbf{v}_{n}$: $\mbox{div}\,\mathbf{v}_{n}\neq 0$. In the two-dimensional
incompressible Euler hydrodynamics, however, singularity formation is
impossible in a \textit{finite} time \cite{wolibner}; possibly, it
can occur in an infinite time. Numerical experiments
\cite{KNNR-07, KNNR-10} yield an exponential increase in vorticity
gradients with the characteristic increment of the order of the maximum
value of vorticity rather than double-exponential growth, as it follows
from the estimates made in \cite{RS78}. The same behavior is observed
in our numerical experiments, which ensure a better resolution (see
the next paragraph).

Let us briefly describe the numerical scheme used.
The equation for vorticity (\ref{vort}) was solved numerically
in a square box with periodic boundary conditions along both coordinates.
The box size was chosen to be equal to unity. To eliminate bottleneck
instability and ensure a sink of energy at small scales,  we introduced
dissipation into the right hand side of Eq. (\ref{vort}), which was taken in
the form of hyperviscosity
\[
(-1)^{n+1}\mu_{n} \nabla^{2n} \omega,\,\, \mu_n =
10^{-20}\left(\frac{2048}{K} \right)^{2n},\quad n=3.
\]
The use of hyperviscosity allows appreciable reduction of the range of
scales where the influence of dissipation is significant and, therefore,
expansion of the inertial region of the spectrum \cite{borue}.
When the spatial resolution was changed, the hyperviscosity coefficient
was scaled in such a manner that the value of dissipation for the shortest
wave perturbations that could be presented on the computational grid
remained constant. The effect of hyperviscosity on the flow as a whole
always remained extremely small: in the numerical experiments, the total
energy was preserved within 0.01 \% .

Equation (\ref{vort}) with hyperviscosity was solved by the pseudo-spectral
Fourier method \cite{canuto}, i.e., the derivatives were calculated and
the Poisson equation (for finding the stream function from vorticity) was
solved in the spectral space, whereas the nonlinear terms were calculated on
a computational grid in the physical space. The transformations from the
physical to the spectral space and back were performed through the Fast
Fourier Transform (FFT) with the FFTW library used for that purpose
\cite{fftw}.
\begin{figure*}[t]
\label{fig1}
\includegraphics[width=0.45\textwidth]{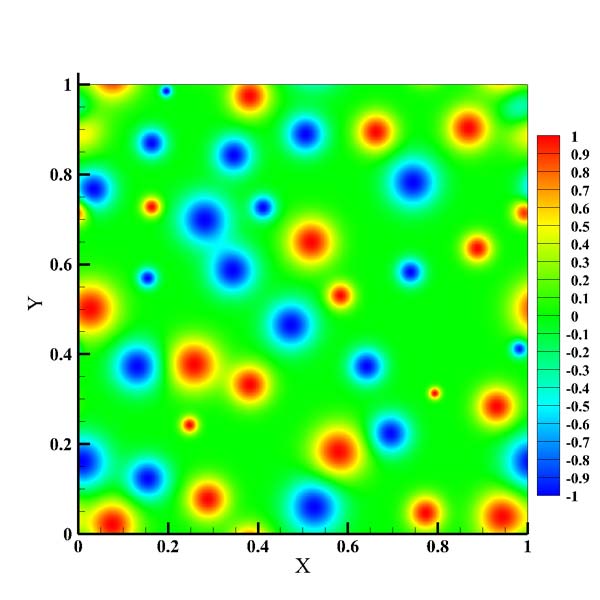}
\hfill
\includegraphics[width=0.45\textwidth]{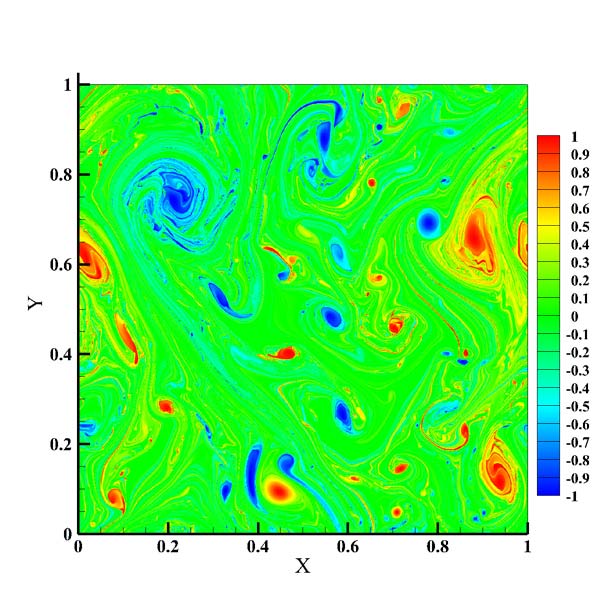}
\parbox[t]{0.45\textwidth}
{\caption{Initial distribution of vorticity, $N=20$}}
\hfill
\parbox[t]{0.45\textwidth}
{\caption{ Distribution of vorticity at $t=100$, $N=20$}}
\label{fig2}
\end{figure*}
The use of hyperviscosity gives rise to a stiff term in the equation;
because of this term, the time step of integration is restricted
for any explicit scheme by a prohibitively small value
($\sim 1/K^{2n}$). To avoid it, time integration was
performed by a hybrid third-order Runge--Kutta / Crank--Nicholson
method \cite{spalart}. The convective term is approximated
explicitly, and an implicit scheme is used for the dissipative
(linear) term.

The computations were performed both on a conventional multiprocessor
cluster (for which the code was parallelized with the help of the MPI library;
up to 128 processors were used) and on a GPU cluster (with the use of
the NVIDIA CUDA technology) at the Computer Center of the Novosibirsk
State University. The spatial resolution was up to 8192$\times$8192
points.

\vspace{0.5cm}{\bf 3. Numerical experiments.}
The initial condition was chosen in the form of two sets of vortices with
positive and negative vorticity. All vortices had a Gaussian shape
with the maximum value of $|\Omega |$ equal to unity and with random sizes
(the vortex radii were uniformly distributed in a certain interval);
the vortex centers were also randomly distributed over the entire domain.
The total vorticity was equal to zero. The number of positive vortices
was equal to the number of negative vortices; this number varied depending
on the spatial resolution from $N=10$ at 2048$\times $2048 to $N=40$
at the maximum spatial resolution.  Variations of the initial conditions
did not change the qualitative behavior of the vortices and their
statistical characteristics. A typical initial condition is shown
in Fig. 1.

Figure~2 shows the vorticity distribution at $t=100$. This time is close
to the maximum value $t_{max}$ at which the computation was terminated.
The time $t=t_{max}$ is determined from the condition that the edge of the
energy spectrum $k_{bound}$ reaches $2/3$ of the maximum value $k_{max}$
in the box. At $t>t_{max}$, the aliasing phenomenon arises \cite{canuto},
where the spectrum is distorted because harmonics located at a distance of
$2k_{max}$ from each other in the spectral space cannot be distinguished on
the computational grid.

As is seen from Fig.~2 (resolution $8192^{2}$), the structure of vortices
at $t=100$ is substantially changed owing to interaction of vortices,
emergence of shear flows generated by vortices, and their stretching.
Figure 3 shows the angularly averaged compensated spectrum of turbulence
$k^{3}E(k)$ at different instants of time. At times close to
$t_{max}$, a plateau is formed in the compensated spectrum, i.e.,
a Kraichnan-type spectrum $E(k)\sim k^{-3}$ is formed almost on two decades.

As was noted previously (see also \cite{KNNR-07,KNNR-10}), such a
spectrum is formed owing to the sharp vorticity gradients.  Figure 4
shows the distribution of $|\mathbf{B}|\equiv|\nabla\Omega|$ at $t = 100$,
which demonstrates a significant increase ($\sim 100$) in the initial
gradients in the vicinity of lines corresponding to the positions
of vorticity shocks. Between the lines, $|\mathbf{B}|$ practically does
not increase. At greater times ($t > 100$), the lines become considerably
longer; correspondingly, the pattern contains more lines, and the
distribution of these lines becomes rather twisted (turbulent).

At the initial stage, at several turnover times $n$
($T_{\ast }=n2\pi/\Omega _{max}$), $|\mathbf{B}|$ grows almost exponentially,
and then the saturation is observed. Thus, for Fig. 5,  the number of turnovers
is $n\approx 4$, and $|\mathbf{B}|$ increases almost by a factor of $300$.
\begin{figure}[t]
\centerline{
\includegraphics[width=0.45\textwidth]{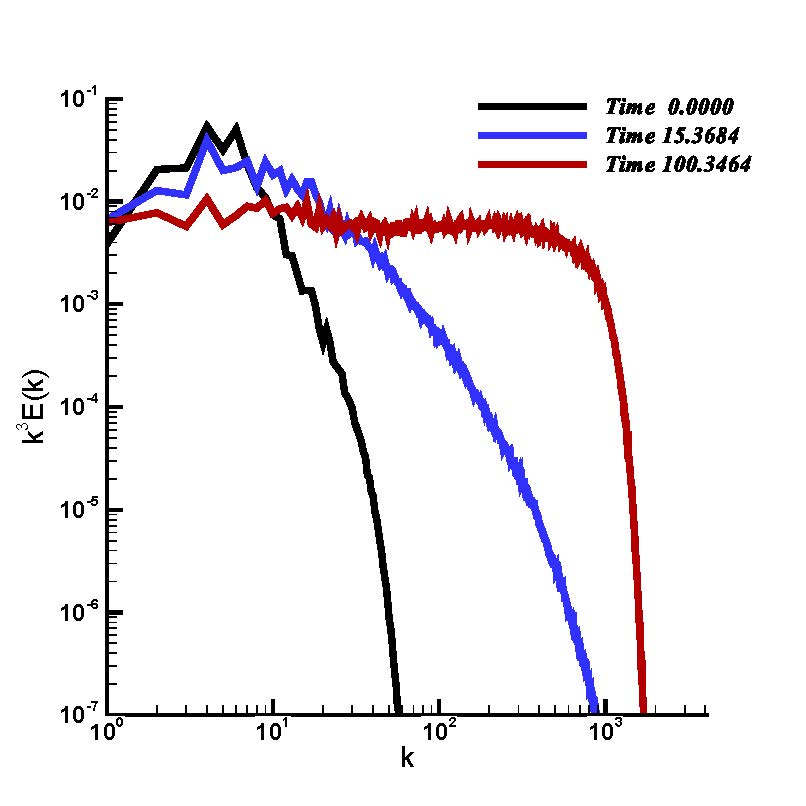}
}
\caption{Compensated spectrum $k^{3}E(k)$ at different
times for the initial distribution shown in Fig. 1}
\label{fig3}
\end{figure}

A significant increase in vorticity gradients and in the density of
lines corresponding to the positions of vorticity shocks testifies that
their role in spectrum formation is rather important. This (principal)
issue was not completely clarified in \cite{KNNR-07,KNNR-10}. To verify numerically
that the Kraichnan-type spectrum $E(k)\sim k^{-3}$ is generated by sharp
gradients of vorticity, we performed spatial filtration, in contrast to
\cite{KNNR-07, KNNR-10}, where filtration in the $k$-space was performed.
In the distribution of the field $\mathbf{B}(\mathbf{r})$, we retained only
those regions where $|\mathbf{B}|$ exceeded some threshold value $B_{0}>0$.
In regions with $|\mathbf{B}|<B_{0}$, the field $\mathbf{B}$ was set to zero.
Using the ``filtered'' field $\widetilde{\mathbf{B}}\left(\mathbf{r}\right) $,
we found the Fourier transform $\widetilde{\mathbf{B}}_{k}$. Further
the Fourier transform of the filtered velocity was determined:
\[
\mathbf{v}_{k}=-\frac{1}{k^{4}}\left[ \mathbf{k\times }\left[
\mathbf{k\times }\widetilde{\mathbf{B}}_{k}\right] \right],
\]
and then the ``filtered'' spectrum $\widetilde{E}(k)$.
(It is evident that this procedure introduces its own contribution to the
spectrum $\widetilde{E}(k)$; this contribution is ensured by the shocks
$|\widetilde{\mathbf{B}}|$ generated by power-law tails decreasing at least
as $k^{-4}$. In other words, filtration alters only the long-wave part
of the spectrum.)
\begin{figure}[t]
\centerline{
\includegraphics[width=0.45\textwidth]{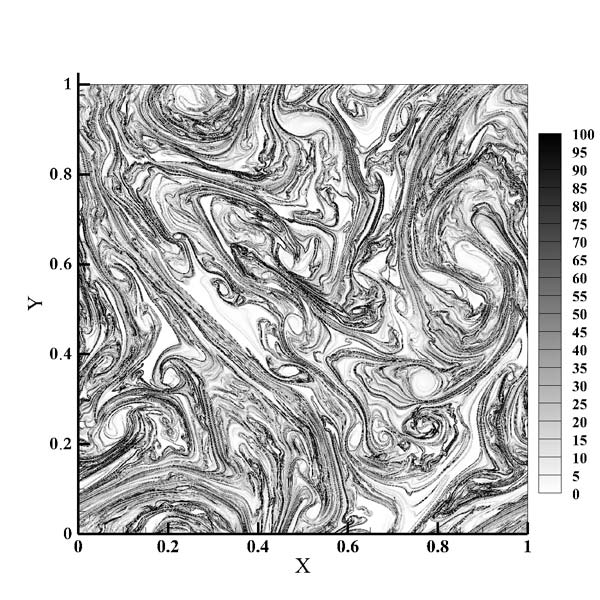}
}
\caption{Distribution of $|\mathbf{B}|$ at $t=100$ for the
initial condition  shown in Fig.~1}
\label{fig4}
\end{figure}
\begin{figure}[t]
\centerline{
\includegraphics[width=0.45\textwidth]{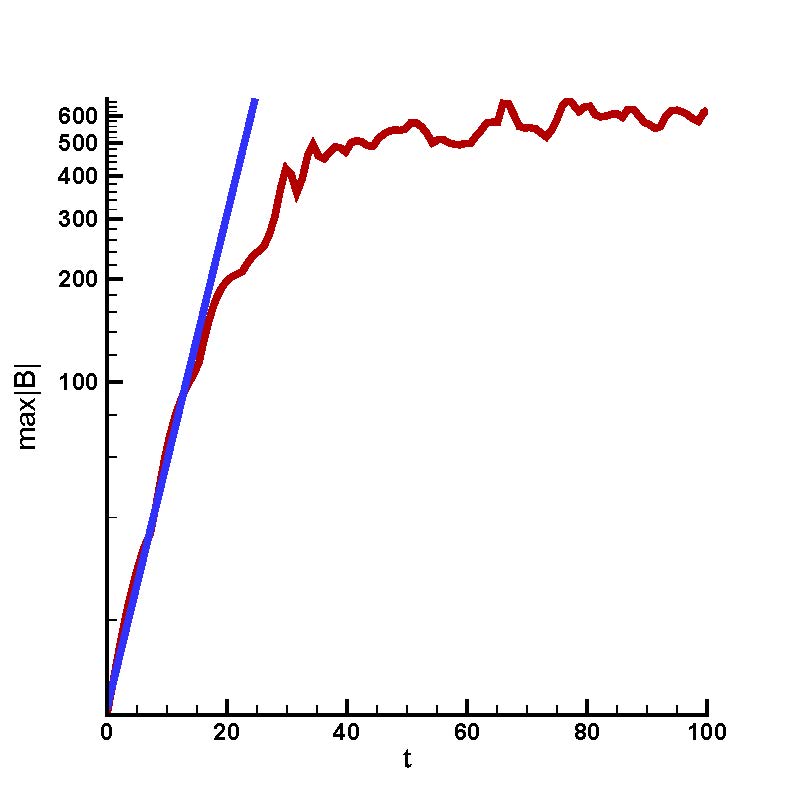}
}
\caption{Maximum value of $|\mathbf{B}|$ versus time for the
initial condition shown in Fig.~1 (logarithmic scale, the straight line corresponds
to the exponential growth)}
\label{fig5}
\end{figure}
Figure~6 shows the ``filtered'' spectra $k^{3}\widetilde{E}(k)$ for
different threshold values $B_{0}$. As a result of such filtration, the
(compensated) spectrum $k^{3}\widetilde{E}(k)$ experienced changes only
in the region of small values of $k$ and remained unchanged in the
region of large values of $k$, which is a numerical consequence of the
fact that the Kraichnan-type spectra are induced by vorticity
quasi-singularities.

The reason for the emergence of the dependence $k^{-3}$ for the spectrum
is closely connected with its angular distribution. Figure 7 shows the
distribution of the two-dimensional spectrum $\epsilon (k_{x},k_{y})$.
This distribution considerably varies with the angle. It is
a number of jets with weak and/or strong overlapping. This behavior is
also clearly visible in the angular dependence of the spectra with
different values of $k$ (Fig.~8). Along each jet, the spectrum
$\epsilon_{k}$ changes as $k^{-4}$, which yields (after
angular averaging) a Kraichnan-type distribution $E(k)$ (\ref{Kraichnan}).

Apart from the spectrum, important characteristics of turbulence
are structure functions.
As was noted above, at times $t$ close to
$t_{max}$, the distribution of the lines of the maximum values of
$|\mathbf{B}|$ becomes substantially more complicated; we can say that
it becomes turbulized.
In particular, the angular overlapping of
jets becomes more noticeable, and the spectrum becomes more isotropic.
To elucidate whether the turbulence considered here is close to
 the Kraichnan-type turbulence, we performed numerical experiments
at $t$ close to $t_{max}$ and obtained the dependence of the
correlation function
\[
D(R)=\langle (\Omega(\mathbf{r})-\Omega(\mathbf{r^{\prime }}))^2
\left([\mathbf{v}(\mathbf{r})-\mathbf{v}(\mathbf{r^{\prime }})]\cdot [\mathbf{R}]\right)/R\rangle
\]
on $R=|\mathbf{r}-\mathbf{r^{\prime }}|$. For the direct cascade, this
dependence on $R$, as was predicted by Kraichnan (see, e.g., review
\cite{boffetta}), should be linear with a proportionality
coefficient $\eta$, which is the flux of enstrophy toward the
region of small scales. Figure 9 shows that this dependence on $R$
is indeed close to linear, especially at small values of $R$.
\begin{figure}[t]
\centerline{
\includegraphics[width=0.45\textwidth]{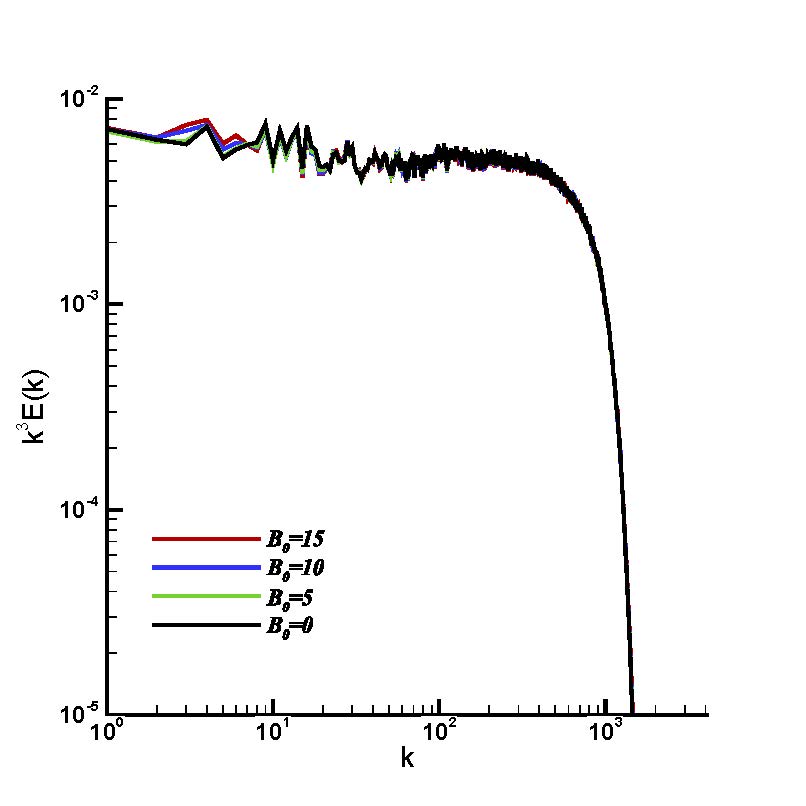}
}
\caption{Filtered compensated spectra $k^{3}\widetilde{E}(k)$
for different threshold values $B_{0}$}
\label{fig6}
\end{figure}
\begin{figure}[t]
\centerline{
\includegraphics[width=0.45\textwidth]{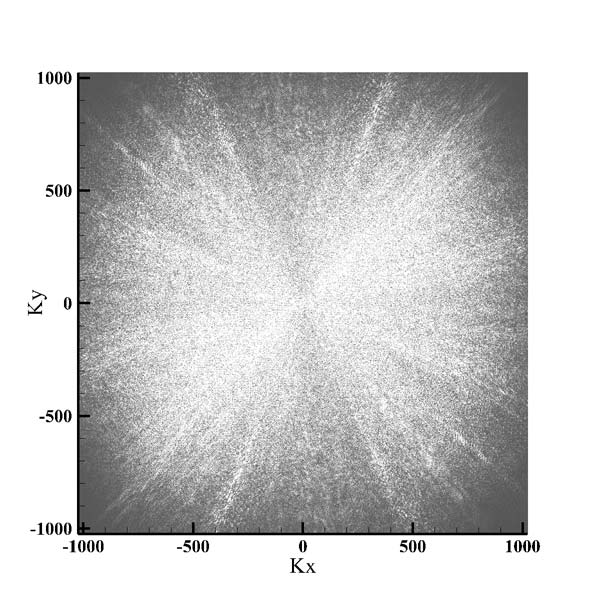}
}
\caption{Distribution of the two-dimensional spectrum
$k^4\epsilon (k_{x},k_{y})$.}
\label{fig7}
\end{figure}
\begin{figure*}[t]
\centerline{
\includegraphics[width=0.3\textwidth]{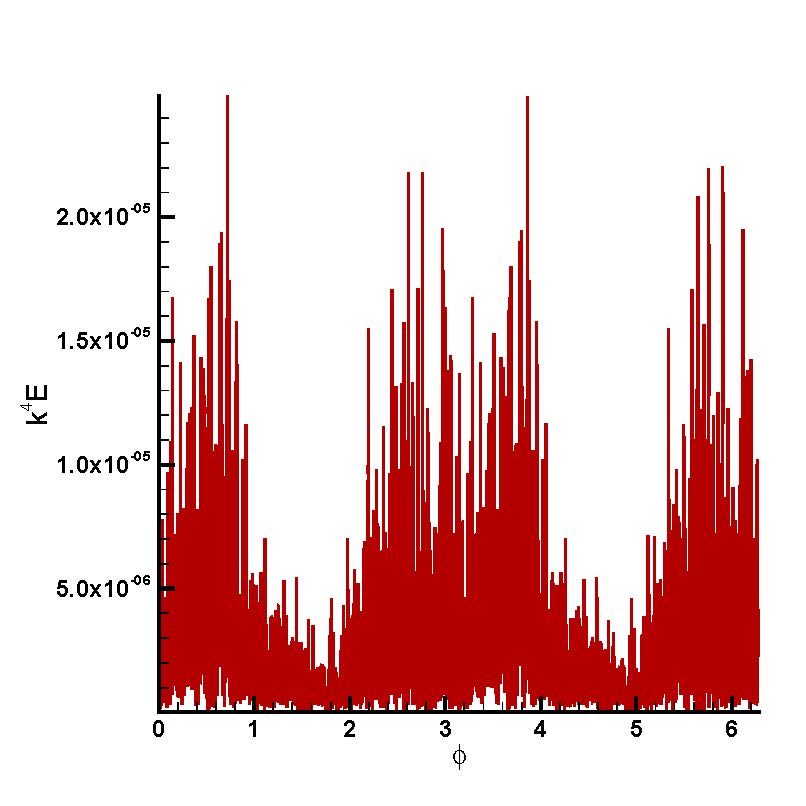}
\includegraphics[width=0.3\textwidth]{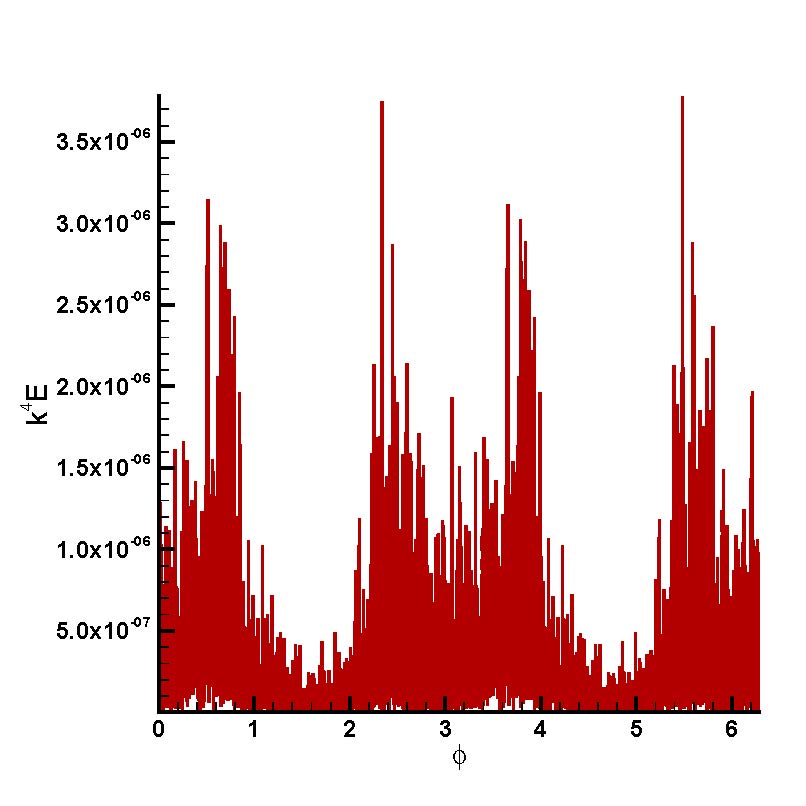}
\includegraphics[width=0.3\textwidth]{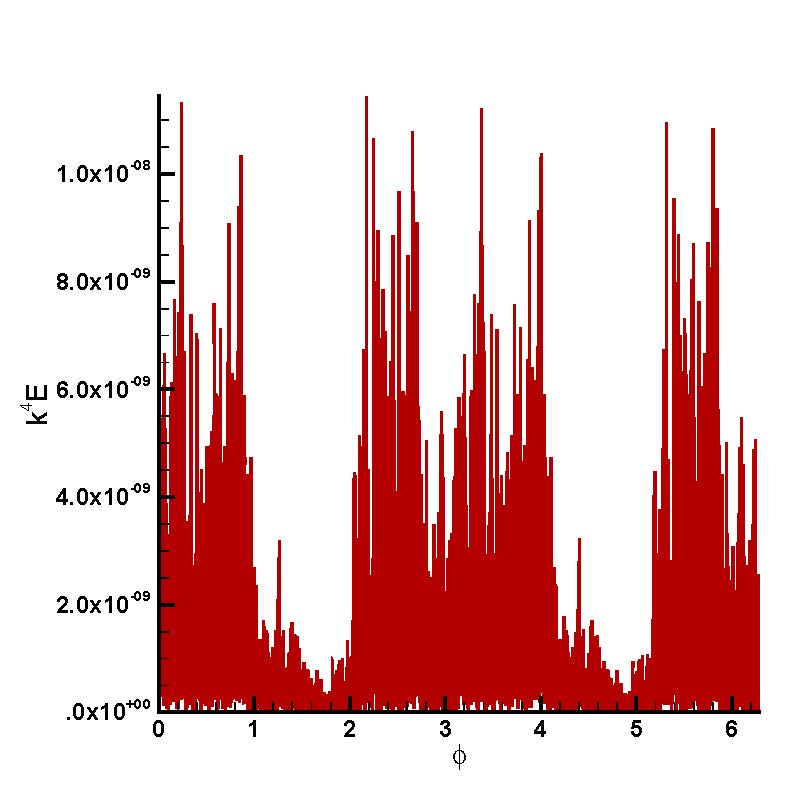}
}
\caption{Compensated spectrum $k^4\protect\epsilon$
versus the angle $\phi$ for $k_0=500$, $k_0=1000$, and $k_0=1500$
}
\label{fig8}
\end{figure*}
At large values of $R$, there is a small inflection, which, in our opinion,
is associated with deviation of the spectrum from the isotropic one,
i.e., with the jet-like structure of the spectrum.
\begin{figure}[t]
\centerline{
\includegraphics[width=0.45\textwidth]{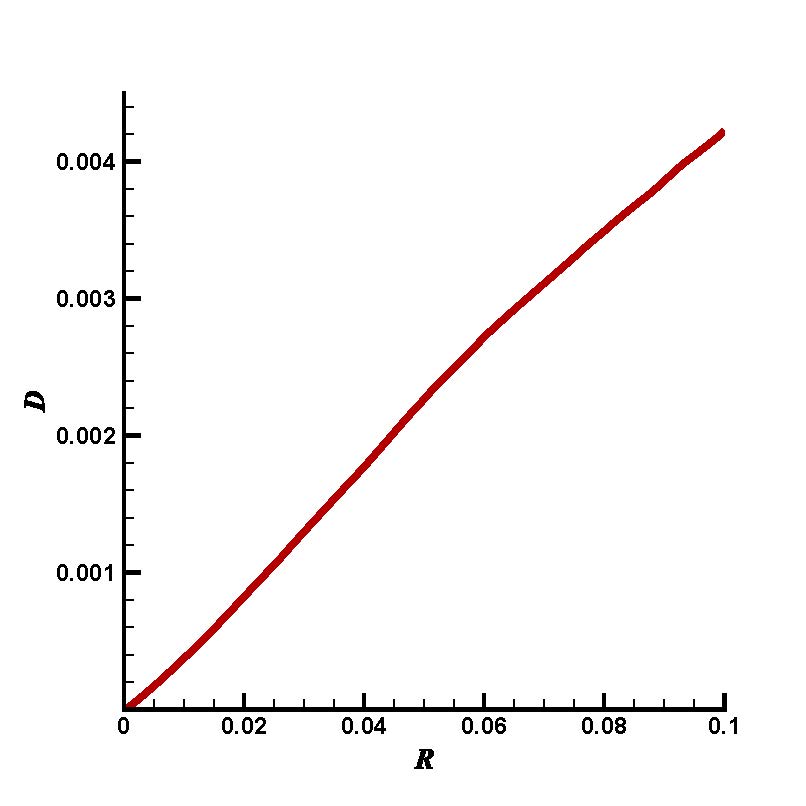}
}
\caption{ Correlation function $D(R)$}
\label{fig9}
\end{figure}

In addition to the correlation function $D(R)$, we also calculated the
structural functions of velocity
$S_n(R)=\left\langle\left[\left(\mathbf{v}(\mathbf{r^{\prime
}})-\mathbf{v}(\mathbf{r})\right)\cdot \frac{\mathbf{r}^{\prime
}-\mathbf{r}}{R}\right]^n\right\rangle$.

It is known that the dependence of the power of the structure functions
$\zeta_n$ ($S_n(R)\sim R^{\zeta_n}$) on $n$ characterizes the degree of
turbulence intermittency. For the Kolmogorov--Kraichnan regime, $\zeta_n$
is a linear dependence of $n$. Our numerical experiment shows that the local
dependence $\zeta_n$ (Fig.~10) deviates from a linear dependence and has
a tendency to saturation, especially at large values of $R$, which testifies to
intermittency of this type of turbulence. It should be noted that all $\zeta_n$ in
the absence of the dissipative scale depend linearly on $n$ as $R\to 0$ (see
Fig. 10). It is caused by the analytical dependence of velocity difference
at small scales.

\vspace{0.5cm}{\bf 4. Conclusion.} The main resume of this work is the fact that the Kraichnan-type power-law
spectrum is formed due to quasi-singularities, which appear in solving
the Cauchy problem for the two-dimensional Euler equation. Though the
collapse as the process of singularity formation in a finite time is
forbidden in two-dimensional hydrodynamics, but there is a tendency to
the emergence of such  singularities. As it follows from our numerical
experiments, vorticity gradients increase by more than two orders of magnitude.
In our opinion, this process can be enhanced in the presence of forcing.
This assumption is based on the presence of a logarithmic correction in terms
of $k$ in the Kraichnan spectrum, which is associated with a weak interaction nonlocality.
It is also of interest that the third-order correlation function
$D(R)$ demonstrates the Kraichnan-type behavior at the stage of turbulence
decay, when a plateau is formed in the compensated spectrum
$k^3E(k)$, which does not contradict the conclusion that spectrum
(\ref{Kraichnan}) is generated by vorticity quasi-shocks. At a sufficiently
late stage of turbulence development, however, the turbulence spectrum has
a rather anisotropic distribution over the angle owing to jets.

\vspace{0.5cm}

The authors  thank G.E. Falkovich and J.J. Rasmussen for useful discussions.
This work was supported by the grant of the Government of the Russian Federation
for state support of academic research performed under supervision of
leading scientists in Russian educational institutions of higher professional
education (contract No.~11.G34.31.0035 dated November 25, 2010 between the Ministry
of Education and Sciences of the Russian Federation, Novosibirsk State
University, and leading scientist), by the RFBR (grant No.~12-01-00943), by the program  "Fundamental problems of nonlinear
dynamics in mathematical and physical sciences" of the RAS Presidium, and by the grant No.~NSh~6170.2012.2 for state support of leading scientific schools of the Russian Federation.
\begin{figure}[t]
\centerline{
\includegraphics[width=0.45\textwidth]{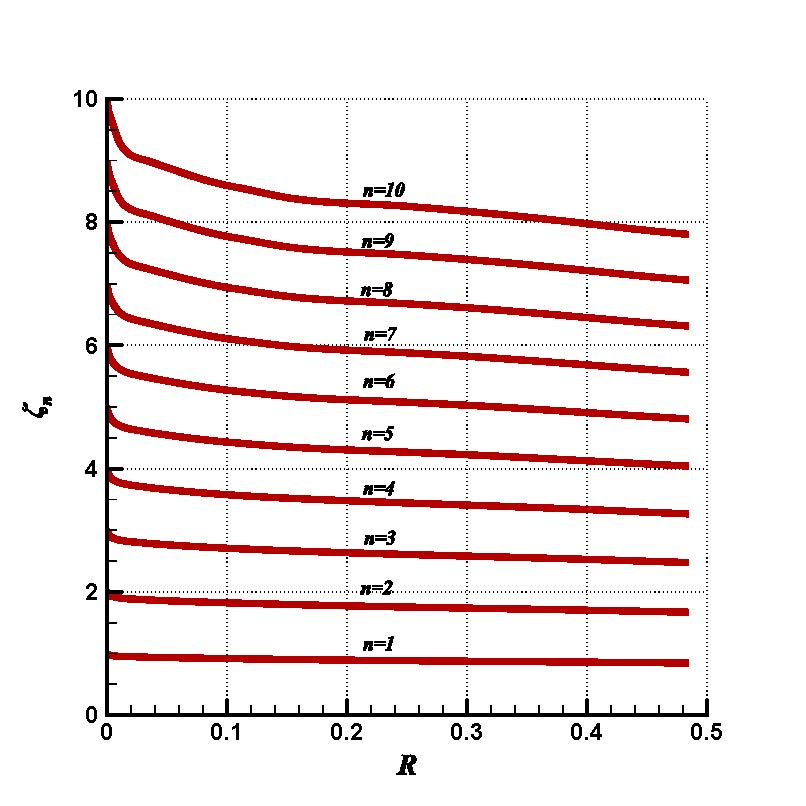}
}
\caption{Exponents $\zeta_n$ (local) as functions of $R$}
\label{fig10}
\end{figure}
E.K. would like to acknowledge his gratitude for hospitality of
the Observatoire de la C\^ote d'Azur (Nice, France), where this paper
was finalized.

\end{document}